\documentclass[twocolumn,english,floatfix,prl]{revtex4}
\usepackage{graphicx}

\makeatletter

\usepackage{slashbox}
\usepackage{hhline}
\usepackage{psfrag}

\usepackage{babel}
\makeatother
\begin{document}

\title{Persistence and Memory in Patchwork Dynamics for Glassy Models}

\author{Creighton K. Thomas,$^{1}$ Olivia L. White,$^{2}$ and A. Alan Middleton$^{1}$}

\affiliation{$^{1}$Department of Physics, Syracuse University, Syracuse, NY 13244,
USA}

\affiliation{$^{2}$Physics Department, Massachusetts Institute of Technology,
Cambridge, MA 02139}

\begin{abstract}
Slow dynamics in disordered materials prohibits direct simulation of their rich nonequilibrium behavior at large scales. ``Patchwork dynamics'' is introduced to mimic relaxation over a very broad range of time scales by equilibrating or optimizing
directly on successive length scales. This dynamics is used to study coarsening
and to replicate memory effects for spin glasses and random ferromagnets. It is also used to find, with
high confidence, exact ground states in large or toroidal samples.
\end{abstract}
\maketitle
The term ``spin glass'' refers both to experimental disordered
magnetic systems
and to theoretical models with enough randomness and frustration in
their interactions to preclude conventional magnetic order.
The experimental systems exhibit a complex cluster of history-dependent
nonequilibrium effects \cite{SGReviews,Vincent}. For example, while a spin glass is ``aged''
at fixed temperature, its magnetic susceptibility slowly changes,
even after waits 20 orders of magnitude longer than the time for single
spin reorientation. Upon further cooling, the material ``rejuvenates'':
its susceptibility reverts to what it would have been without the
wait. But amazingly, the system does retain a ``memory'' of its
history and when temperature returns to that at which aging took place,
susceptibility nears its aged value. In fact, waits at multiple temperatures
can be stored and recovered in \cite{MiyashitaVincent}.
Similar effects are seen in a variety of experimental systems and
multiple explanations have been proposed \cite{Vincent,ARM}. Yet despite
thirty years of study, the nature of spin-glass dynamics -- and of
aging and memory effects in other ``glassy'' materials -- remains
controversial and ill-understood. In particular, how these effects
are related to the temporal evolution of correlations
is an open question. 

In this paper, we present ``patchwork dynamics'', a numerical
approach for studying growth of correlations and non-equilibrium effects
over a wide range of length and time scales in systems with quenched
disorder. Patchwork dynamics proceeds by a succession of coarse-grained
equilibrations -- or optimizations at zero temperature -- and provides
a framework for investigating the relation between the evolution
of microscopic correlations and the complex nonequilibrium effects
observed in experimental spin glasses. This approach therefore replaces
dependence on time by dependence on length scales.
It can be used to study coarsening and
the persistence of the initial state, to replicate
memory and rejuvenation effects, to visualize how disordered
systems store their history, and also as a ground state algorithm
for systems that are otherwise difficult to optimize. 

As an initial application, in this paper we investigate the two-dimensional (2D) Edwards-Anderson Ising spin glass model (ISG) and the 2D random bond ferromagnet (RBFM), both at zero temperature. 
For both models, the Hamiltonian, ${\cal H}$, has the form
${\cal H}=-\sum_{\langle ij\rangle}J_{ij}s_{i}s_{j}$, where the Ising
spin variables $s_{i}=\pm1$ lie on a $d$-dimensional lattice and the
$J_{ij}$ are mean-zero Gaussian random variables for the ISG and are
random, but positive, for the frustration-free RBFM model.
The nature of the low-temperature state in the ISG, including even
the number of states, is still a subject of study
\cite{FisherHuseLambda,pictures} and
the response to perturbations, such as modifications of $T$, the
$J_{ij}$, or boundary conditions (BCs) is complex and only partially
understood. 
Note that the 2D ISG is ``glassy'' only as $T\rightarrow0$, but we use it as a model for
the low-temperature phase of glassy magnets where temperature is
irrelevant. 
Patchwork
dynamics is a general approach, with other immediate applications
including \cite{future} non-equilibrium dynamics in three-dimensional
spin glass models and 2D dimer models at finite temperature.

Dynamics in disordered materials is extremely slow since arbitrarily
large groups of spins must rearrange to explore the low-free-energy
phase space described by ${\cal H}$. A homogeneous ferromagnet
(uniform positive $J_{ij}$) can move from any initial state to its
ground state via energy lowering flips of small numbers of spins.  By
contrast, in a spin glass, arbitrarily large collective spin flips
must occur \cite{kFlip}. It has been argued that the case of the RBFM
is similar: disorder pins domain walls separating constant-spin
domains on arbitrarily large scales
\cite{FisherHuseRBFM}. Furthermore, the time to flip a collection of
spins grows quickly with the number of spins in the cluster. The
associated time scale is $\sim\tau_{0}e^{B/T}$ with microscopic time
scale $\tau_{0}$ and $B$ the energy barrier to the flip. For scale
$\ell$ clusters, the typical barrier $B(\ell)$ will grow with length
scale, since more improbable events must occur simultaneously. The
distribution of barriers will have a broad range of values and thus
the distribution of flip times will be even more broadly
distributed. Thus rough separation of time scales is expected for
geometrically separated $\ell$. In particular, this occurs for the
simplest hypothesized scaling form $B\sim\ell^{\psi},$ with $\psi$ a
model-dependent scaling exponent, and the width of the distribution of
$B$ given by its typical value. (Numerical evidence sometimes has
suggested a power law growth \cite{RiegerAging}, or logarithmic
barriers, and the multiplicity of
barriers also needs to be considered.)  One numerical consequence is
that the range of length scales that single-spin-flip Glauber dynamics
can probe is severely limited by the quick growth of time scales with
length, even though it can give some useful information about the
evolution of correlations in spin glasses and other disordered systems
\cite{RiegerAging,ARM,GhostDomains}.

The dependence of barriers on length scale motivates patchwork dynamics, which
takes advantage of the separation of time scales to mimic the development
of correlations.  It operates over a succession of increasing spatial scales,
$\ell_{m}=1,2,\ldots,2^{m},\ldots$ , starting from an initial spin configuration
on a given sample of fixed size $L$.  For each $m$, we ``equilibrate'' the
sample by optimizing (or equilibrating) randomly-chosen subsystems -- patches
-- of size $\ell_{m}$, with fixed boundary conditions on a patch given by the
surroundings. 
After applying sufficient patches at scale $\ell_{m}$, the scale
is increased to the next in the sequence, $\ell_{m+1}$, and the procedure is
repeated.  (Results are similar for arithmetical sequences of $\ell_m$.) We
implement two versions of this dynamics. In the ``complete'' version, we repeat
the optimization at each scale until no more improvements are possible. In the
``mean coverage'' approach we apply $c(L/\ell_{m})^{d}$ patches, covering the
system $c$ times at that scale. In both versions, patch placement is
independent of the actual barriers, so the dynamics cannot replicate the
fine-grained evolution of correlations.  Refinements might include
using a barrier-dependent success probability in placing patches or
determining (non-square) patches that have the lowest estimated
barrier \cite{BarriersByOpt}. Note that the general procedure is numerically
tractable only when equilibration or optimization is fast enough at each length
scale \cite{OptAlg}. 

In spirit, patchwork dynamics resembles simulations
used to study the mosaic picture of structural glasses' dynamics, where
atoms are equilibrated inside of a fixed shell, and analytic work on
kinetically constrained models of glasses \cite{MosaicKCM}. Multiscale approaches have also been used to study spin glasses on hierarchical
lattices \cite{HierarchicalPatches} as well as in optimization
\cite{RenormGenetic}.  However, here we focus primarily on using
equilibrium or ground state solutions in the study of non-equilibrium
dynamics.

We apply patchwork dynamics to studying equilibration of the 2D ISG
Gaussian model and 2D RBFM model (with uniform distribution for
$J_{ij}\in[0,1]$) on square lattices with toroidal boundary
conditions. While we describe and carry out simulations for the case
of optimization, equilibration on subsamples can replace ground states
throughout. Calculations were carried out for systems up to size
$256^{2}$ with patch sizes up to $\ell_{m}=128$, using on the order of
$4\times 10^3$ samples. The mean-coverage approach with $c=150$ is
nearly indistinguishable from the complete approach and qualitative
behaviors, including power law exponents for large $\ell_{m}$, are
independent of $c\ge 8$. Furthermore, the number of spins flipped at
each scale approaches a $c$-independent limit, consistent with the
assumption that changes over geometrically separated length scales
occur on well-separated time scales.

\begin{figure}
\includegraphics[width=1\columnwidth]{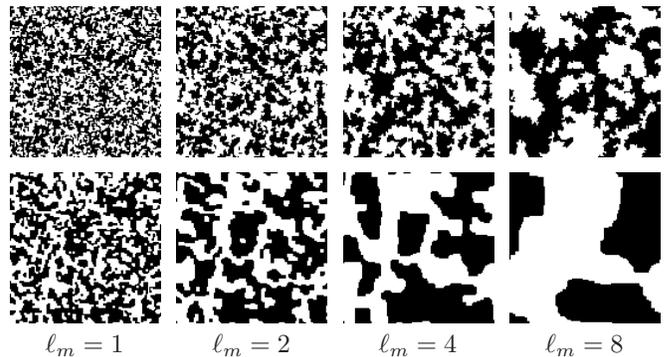}
\caption{\label{cap:Sequences}Domain growth under patchwork dynamics for
a $L^{2}=128^{2}$ sample. Light spins are aligned with one of the
ground states, while dark spins are aligned with the other. The initial
spins are random. At each scale $\ell_{m}$, randomly
selected patches of dimension $\ell_{m}\times\ell_{m}$ are optimized,
with fixed spins exterior to the patch, until a stable configuration
is reached. The upper row shows a history for the Ising spin glass,
while the lower row is for the random bond ferromagnet, for which
the typical domain size $b(\ell_{m})$ grows faster than $\ell_{m}$.}
\end{figure}

\begin{figure}
\includegraphics[width=1\columnwidth]{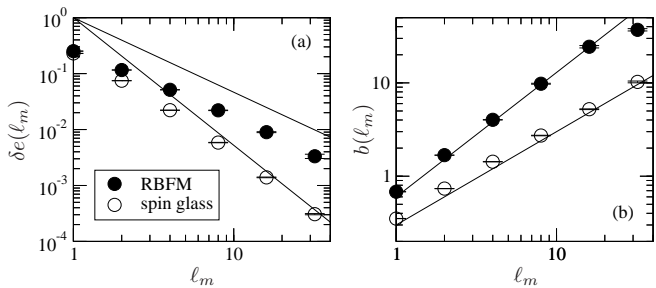}
\caption{\label{cap:Eandb}Plots of (a) the residual energy density $\delta e(\ell_{m})$
above the ground state energy density and (b) the domain scale $b(\ell_{m})$,
as a function of the patch size $\ell_{m}$, for the 2D Ising spin glass and
the random bond ferromagnet. Straight lines indicate power law
behaviors described in the text.
}
\end{figure}

To study the approach to the final state, at each scale $\ell_{m}$ we
compare our solutions with the doubly degenerate optimal $T=0$
solution.  For the RBFM, this configuration has uniform spin. For the
ISG, we compare to the extended ground state
\cite{ThomasMiddletonXGS}, the optimal combination of spin variables
and choice of periodic/anti-periodic boundary conditions minimizing
the EA Hamiltonian, and we set boundary conditions
accordingly. 
We estimate a coarsening length scale
$b(\ell_{m})$, as evident in Fig.~\ref{cap:Sequences},
by the mean Manhattan distance from a randomly chosen
point to a domain wall. We also compute the residual energy density
$\delta e(\ell_{m})=\overline{L^{-d}[E(\ell_{m})-E_{\mathrm{GS}}]}$,
where the overline indicates an average over disorder, $E(\ell_{m})$
is the total Hamiltonian at the completion of the computations for
stage $m$, and $E_{\mathrm{GS}}$ is the ground state energy for the
sample. Fig.~\ref{cap:Eandb} shows residual energy, $\delta e$, and
domain size, $b$, as a function of patch scale, $\ell_{m}$.  The spin
glass results are consistent with $\delta e\sim\ell_{m}^{\theta-2}$,
with $\theta=-0.27$ \cite{Theta}, $b\sim\ell_m$. The results for the 2D RBFM are
consistent with the expectation \cite{FisherHuseRBFM}
$\delta e\sim\ell_{m}^{-4/3}$ and
$b\sim\ell_m^{4/3}$. The accuracy of the slopes is about $\pm 0.1$ in
each case, with the expected power laws indicated by straight lines in
Fig.~\ref{cap:Eandb}.

In a spin glass, the decay of magnetization $M$
from a fully magnetized initial condition is closely related
to the persistence of a random initial spin configuration.
Given a random initial condition, let $p(\ell_{m})$ be the probability
a spin points in its $t=0$ direction. For a spin glass, the fully
magnetized initial condition is random with respect to the
bonds, so $M(\ell_{m})=2p(\ell_{m})-1$. Motivated by other known cases \cite{BrayCoarsen}, Fisher and Huse \cite{FisherHuseLambda}
conjecture that in the 3D EA model,
$2p(\ell_{m})-1\sim\ell_m^{-\lambda}$,
where $\lambda$ is an independent exponent describing the dynamics.
A second exponent describing memory decay is the persistence exponent $\theta'$
\cite{BrayCoarsen,Persistence}, where the 
probability a spin never has flipped is conjectured to decay as $P(\ell_m)\sim \ell_m^{-\theta'}$.

\begin{figure}
\includegraphics[width=1\columnwidth]{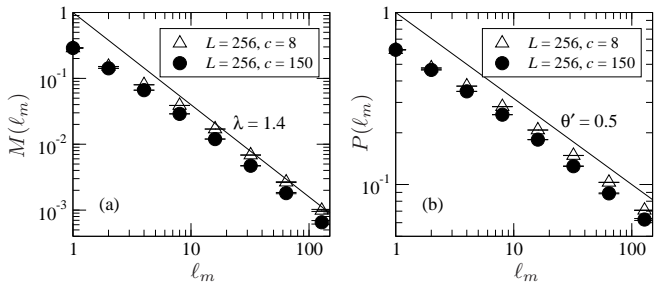}
\caption{\label{cap:persistence}Plot showing decay of the remanent
magnetization and the persistence of the initial spin configuration in
the Ising spin glass. The dependence of (a) magnetization $M(\ell_m)$ and
(b) unflipped spins $P(\ell_m)$ are described by power laws with slopes of
$-\lambda=-1.4\pm0.1$ and $-\theta'=-0.5\pm 0.05$, respectively.}
\end{figure}

We use patchwork dynamics to investigate the values of $\lambda$ and
$\theta'$ in the ISG (Fig.~\ref{cap:persistence}). For $\ell_m>16$, a
power law decay describes the data well with $\lambda_{\mathrm{SG}}=1.4\pm0.1$ and
$\theta'_{\mathrm{SG}}=0.5\pm0.05$.
Estimates of systematic rather than statistical error dominate total error. 
Details of the sequence of patch size do not affect our estimate of $\lambda$, as we have checked by applying patches at every scale,
$\lambda=1,2,3,\ldots,L/2$. Furthermore, the effective
exponents (slopes on log-log plots) are nearly independent of $c$
for $L\ge 16$ and $8<c<150$. 
Similarly for the RBFM, we find $\lambda_{\mathrm{RB}}=1.4\pm0.1$.
Note that under patchwork dynamics, the boundaries of overlapping patches store
memory since the dynamics optimizes (or equilibrates) all spins except
those on the boundaries.

The dramatic rejuvenation and memory effects seen in temperature cycling experiments on spin glasses must arise from a special persistence \cite{Vincent}.  Memory of magnetic susceptibility at a given temperature
indicates that the pattern of magnetization must be encoded into the
spin configuration and realized in physical space via domain growth
at distinct temperatures. Similar memory effects are observed when the strength of exchange interactions are perturbed rather than the temperature \cite{NumericsHard,ARM,GhostDomains}. 
It is likely that this ``disorder chaos'' is related to ``temperature chaos'' and memory in experimental spin glasses.
How disordered media like spin glasses can retain spin information so effectively is a central issue in understanding their out of equilibrium behavior.

We use patchwork dynamics to investigate memory effects in the 2D ISG.  Since there is no finite-temperature spin glass phase, between ``equilibrations'' we perturb exchange interactions. We start with random initial conditions and couplings $J_{ij}$.
First, we apply patches up to size $\ell_{\mathrm{max}}^{(1)}$.  
Second, we perturb bond strengths by an amount $\Delta$, via $J_{ij}\rightarrow
J_{ij}'=\frac{J_{ij}+\Delta K_{ij}}{\sqrt{1+\Delta^{2}}}$, with
$K_{ij}$ independent Gaussian random variables, and then we
apply patches up to size $\ell_{\mathrm{max}}^{(2)}$. 
Finally, in the third stage, we revert to the $J_{ij}$ and age using
patches up to size $\ell_{\mathrm{max}}^{(3)}$.
After steps $k=1,2,3$, system configuration is denoted by $s_i(k)$.
The two ground states with couplings $J_{ij}$ and $J_{ij}'$ are correlated on length scales smaller than the ``chaos'' length, $\xi\sim \Delta^{-1/\zeta}$, with
$\zeta=d_s/2-\theta\approx 0.91$ \cite{Chaos}. 
To quantify aging and memory effects, we use the sample-averaged spin overlap
$q(k)=L^{-2}\overline{\sum_i {s_i(1)s_i(k)}}$, for stages
$k=2,3$.

The full parameter space is four-dimensional, defined by
$\ell_{\mathrm{max}}^{(1,2,3)}$ and $\Delta$, so we focus on particular
regions. 
We set $\ell^{(1)}_{\mathrm{max}}=\infty$, so that after the first
stage the system is in the ground state for the original couplings $J_{ij}$,
giving $\ell_{\mathrm{max}}^{(1)}\gg \ell_{\mathrm{max}}^{(2,3)}$.
We then
investigate the two limits of large and small changes in exchange interactions, i.e., large and small $\Delta$.
For all memory simulations, we fix $c=50$.  

\begin{figure}
\includegraphics[width=1\columnwidth]{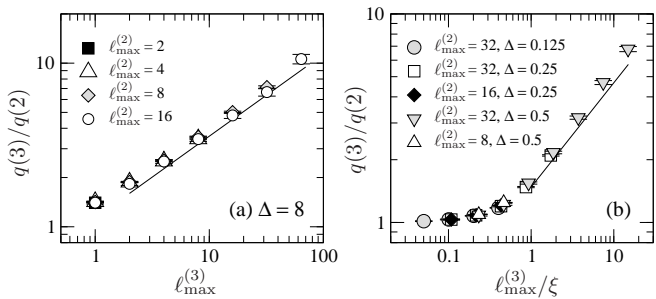}
\caption{\label{cap:Memory}Plots of the strength of memory in the
2D Ising spin glass at zero temperature, for strong and weak
chaos. The first aging stage is long enough to
reach the ground state for a given bond realization $J_{ij}$
(with $L=128$).
The system is coarsened under patchwork dynamics 
up to lengths $\ell_{\mathrm{max}}^{(2,3)}$ in
stages 2 and 3, using bonds $J'_{ij}$ and $J_{ij}$, respectively.  (a) 
For large $\Delta$, $\xi\approx 1.0$ and the final spin overlap
$q(3)$ grows as $q(3)/q(2)\sim [\ell_{\mathrm{max}}^{(3)}]^\kappa$, with
$\kappa=0.5\pm0.05$, for $q(3)$ restricted to $q(3)<0.5$.
(b) For smaller $\Delta$,
using $\xi=2.0(\Delta)^{-1/\zeta}$, and
$\ell_{\mathrm{max}}^{2}>\xi$, $q(3)$ is constant until $\ell_{\mathrm{max}}^{(3)}$
exceeds $\xi$, and $q(3)/q(2)$ collapses to a single function of
$\ell_{\mathrm{max}}^{(3)}/\xi$ (again restricting data to $q(3) < 0.5$). Solid lines
indicate $\kappa=0.5$.}
\end{figure}

When $\Delta$ is large, $\xi = 1$ and the ground state with $J_{ij}'$ is nearly uncorrelated to that for bonds $J_{ij}$.  Thus the spin overlap after the second stage scales as $q(2)\sim[\ell_{\mathrm{max}}^{(2)}]^{-\lambda}$.  
During the third stage, we find that the overlap increases with increasing
$\ell_{\mathrm{max}}^{(3)}$, as the ground state is recovered with the original
$J_{ij}$. 
We find that $q(3)/q(2) \sim [\ell_{\mathrm{max}}^{(3)}]^{\kappa}$, with $\kappa=0.5\pm 0.05$, for $q(3)<0.5$ 
and $8\le \ell_{\mathrm{max}}^{(3)}\le 64$ (for
$q(3)>0.5$, the increase of $q(3)$ is independent of $q(2)$).
Fig.~\ref{cap:Memory}(a) shows results for $\Delta=8$.
In fact, there is no obvious difference in the behavior of $q(3)$ between
when $\ell_{\mathrm{max}}^{(3)}<\ell_{\mathrm{max}}^{(2)}$ and
when $\ell_{\mathrm{max}}^{(3)}>\ell_{\mathrm{max}}^{(2)}$.

We study the crossover from weak to strong chaos by using smaller
values of $\Delta$.  The overlap
length relating ground states for $J_{ij}$
and $J'_{ij}$ is $\xi \approx (2.00\pm 0.15)\Delta^{1/\zeta}$.
For $\ell_{\mathrm{max}}^{(2)}<\xi$, $q(2)$ is near unity.
For $\ell_{\mathrm{max}}^{(2)}>\xi$, memory recovery in the third stage only occurs when
$\ell_{\mathrm{max}}^{(3)}>\xi$.
As shown in Fig.~\ref{cap:Memory}(b), the ratio $q(3)/q(2)$ plotted against 
$\ell_{\mathrm{max}}^{(3)}/\xi$ exhibits a convincing numerical collapse to a
single function, for $q(3)<0.5$. 
The asymptotic memory growth is consistent with
$q(3)/q(2)\sim [\ell_{\mathrm{max}}^{(3)}/\xi]^\kappa$.

Patchwork dynamics at $T=0$ also provides a simple and fast technique
to solve optimization problems quickly \cite{RenormGenetic} in large
systems or in systems on a toroidal lattice, which are otherwise
difficult to study, especially for continuous disorder and arbitrary
boundary conditions \cite{RiegerAging,ThomasMiddletonXGS}.  For
example, on the 2D ISG on a toroidal lattice, using fixed BC patches
of size $L-1$ with coverage $c=16$ and choosing the best evolved state
from 20 random histories (a history is defined by an initial condition
and patch placement sequence), we found the exact extended ground
state in all of $10^{4}$ samples of size $L=256$. We also
found the
exact ground state in $10^3$ samples of size $L=32$ and $L=64$ with
arbitrary boundary conditions, with the same protocol.  The mean time
to find the ground state is approximately $1.5$ histories. Even using
smaller patches of size $L/2$ and $c=50$ in systems of size $L = 128$,
we found the exact extended ground state on a torus using 120
histories with no failures in $10^{4}$ samples; half-size patches can
therefore be used to find the ground state. Such techniques are
currently being studied to compute precise exponent values for the 2D
ISG and to increase the size of 3D systems that can be studied with
high confidence \cite{future}.

In conclusion, coarse-grained evolution on a sequence of length scales mimics the long time scales associated with non-equilibrium evolution in disordered systems, greatly
reducing computational time and also providing a theoretical framework for studying non-equilibrium effects in such systems. We demonstrate the use of this approach in
determining coarsening and persistence exponents, finding exact ground
states, and replicating aging and memory effects.  One important further
application is to extend these investigations to 3D Ising spin glasses
\cite{future}.  Additionally, the use of this dynamics could be coupled with hierarchical approaches for estimating barriers in glassy models \cite{BarriersByOpt}.

This work was supported in part by NSF grant DMR 0606424. We thank
Daniel Fisher for stimulating discussions, Frauke Liers for confirming
our results on exact ground states on toroidal systems, and the Aspen
Center for Physics for its hospitality.

\end{document}